\documentclass[12pt,showpacs,preprint,eqsecnum,amsmath,amssymb]{revtex4}
\usepackage{graphicx}
\usepackage{amsmath}
\usepackage{amssymb}
\begin{document}
\title{S-MATRIX POLES CLOSE TO THRESHOLDS  \\
IN CONFINED GEOMETRIES}
\author{Giorgio Cattapan}
\affiliation{Dipartimento di Fisica ``G. Galilei'', Universit\`a di Padova,
Via F. Marzolo 8, I-35131 Padova, Italy \\
Istituto Nazionale di Fisica Nucleare, Sezione di Padova, 
Via F. Marzolo 8, I-35131 Padova, Italy}
\author{Paolo Lotti}
\affiliation{Istituto Nazionale di Fisica Nucleare, Sezione di Padova, 
Via F. Marzolo 8, I-35131 Padova, Italy \\
Dipartimento di Fisica ``G. Galilei'', Universit\`a di Padova,
Via F. Marzolo 8, I-35131 Padova, Italy}
\begin{abstract}
We have studied the behavior of the $S$--matrix poles
near threshold for quantum waveguides coupled to a cavity with a defect.
We emphasize the occurrence of both dominant and shadow poles
on the various sheets of the energy Riemann surface, and show that
the changes of the total conductivity near threshold as the cavity's width 
changes can be explained in terms of dominant to shadow pole
transitions.
\end{abstract}
\pacs{73.63.Nm \newblock{Quantum wires},73.23.Ad \newblock{Ballistic 
transport},72.10.Fk \newblock{Scattering by point defects, dislocations, 
surfaces, and other imperfections (including Kondo effect)}
\newline
\null
\vfill
\hfill Preprint DFPD/07/TH12}
\maketitle
In multichannel scattering, the poles of the $S$--matrix lie in general
on different sheets of the Riemann energy surface. Poles on the negative
real axis of the physical sheet correspond to bound states of the system,
while poles on the unphysical sheets close to the physical energy axis 
are associated to resonant states. In addition to the resonance, 
{\em dominant} poles, however, there are also poles on unphysical sheets, far 
away from the physical region, which do not have in general observable effects,
and have been referred to as {\em shadow} poles \cite{et64}. The possible
role of shadow poles has been discussed over the years in the context 
of particle \cite{fu75}, nuclear \cite{bj87} and atomic \cite{ht75} physics,
as well as in laser--induced multiphoton processes \cite{ps88,dp90} . Their 
study is of particular relevance when the scattering process depends upon some 
tunable parameter; as this parameter is changed, the $S$-matrix poles move on 
the various sheets of the energy Riemann surface, and may pass a scattering
threshold. In so doing, some shadow pole may approach the physical region, 
thereby becoming dominant, and producing observable effects, whereas a 
previously dominant pole may retire to a less exposed position 
\cite{rn82}. 

\begin{figure}[ht]
\centerline{\includegraphics[width=10 truecm,angle=0]{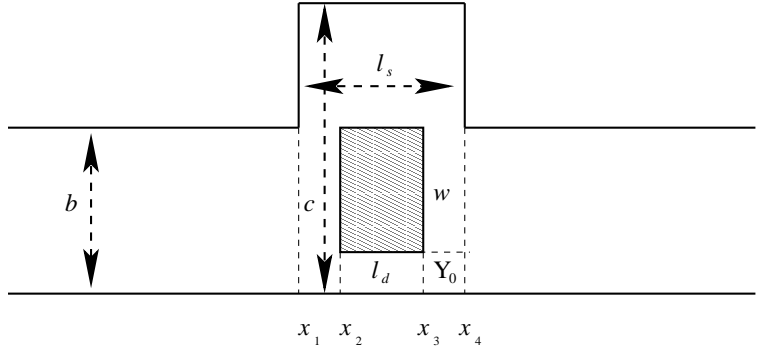}}
\caption{A stubbed quantum waveguide of width $b$ and infinite length, with
a stub of width $c$ and length $l_s$. The stub contains a defect with 
dimensions $w\times l_d$.}
\label{fig1}
\end{figure}

In this paper we would like to point out another situation,
where shadow and dominant poles may exchange their roles, and
give rise to non--trivial observable effects near threshold. It stems
from recent developments in nanotechnology, which allow one to obtain
a strictly two--dimensional electron gas subject to confined geometries
\cite{da95,fg99}. To be definite, we shall consider the device of Fig.
\ref{fig1}, where a resonant cavity or stub having width $c$ and
length $l_s$ is coupled to a uniform guide of indefinite length and
width $b$.  The stub contains a region, depicted by the shaded area
in Fig. \ref{fig1}, with a defect described by a potential field
$V(x, y)$. For high--purity materials and at low temperatures, the
electron's motion inside the duct is ballistic, and can be described as 
a scattering process \cite{da95,fg99}, the conductivity of the quantum 
circuit being expressible in term of the transmission coefficients of the 
system.
       
Recently, we have developed an $S$--matrix approach to stubbed wave
guides with defects, which allows for an accurate numerical solution
of the scattering problem even when some critical dimension of the
system gets large \cite{noi07}. We start from the two--dimensional
Schr\"odinger equation 
\begin{equation}
\left \{ - \frac{\hbar^2}{2 m^{\ast}}\nabla^2_{\scriptscriptstyle{2}} 
+ V(x, y)\right \}\Psi (x, y) = E \Psi (x, y)~~,
\label{scho1}
\end{equation}  
where $\nabla^2_{\scriptscriptstyle{2}}$ is the two--dimensional Laplace
operator, $E$ the total energy, and $m^\ast$ represents the effective
mass of the electron in the conduction band. Assuming hard--wall
boundary conditions, the total wave function $\Psi (x,y)$ is
expanded in terms of the transverse mode eigenfunctions in the lead
and in the cavity, and the Schr\"odinger equation (\ref{scho1}) is
replaced by an (in principle) infinite set of coupled, one--dimensional
Schr\"odinger equations. The latter can be reduced to linear, algebraic
equations matching the wave function and its first derivative at the
various interfaces delimiting the duct from the cavity, and the
empty part of the cavity from the region where the potential acts.
Thus, the scattering operator for each segment in the quantum circuit
can be evaluated through linear algebra. The total $S$--matrix of the
device is finally obtained from the partial scattering operators by 
recursively applying the $\star$-product composition rule, which expresses 
the overall scattering matrix ${\mathbf S}$ in terms of the partial scattering
matrices ${\mathbf S}^{(a)}$ and ${\mathbf S}^{(b)}$ as \cite{da95,noi07}
\begin{equation}
\mathbf{S} = 
\left(
\begin{matrix}
\mathbf{S}_{11}~ & ~\mathbf{S}_{12} \\
\noalign{\vspace{5 pt}}
\mathbf{S}_{21}~ & ~\mathbf{S}_{22}
\end{matrix}
\right) =
\mathbf{S}^{(a)} \bigstar \mathbf{S}^{(b)}~~,
\label{starp}
\end{equation}       
where
\begin{subequations}
\begin{eqnarray}
\mathbf{S}_{11}  = &\mathbf{S}^{(a)}_{11} + \mathbf{S}^{(a)}_{12}
\mathbf{S}^{(b)}_{11}\left(\mathbf{1} - \mathbf{S}^{(a)}_{22}
\mathbf{S}^{(b)}_{11}\right)^{-1}\mathbf{S}^{(a)}_{21}~~, 
\label{str11} \\
\mathbf{S}_{12} = & \phantom{\mathbf{S}^{(a)}_{11} + \mathbf{S}^{(a)}_{12}}
\mathbf{S}^{(a)}_{12} \left(\mathbf{1} - \mathbf{S}^{(b)}_{11}
\mathbf{S}^{(a)}_{22}\right)^{-1}\mathbf{S}^{(b)}_{12}~~, 
\label{str12} \\
\mathbf{S}_{21}  = & \phantom{\mathbf{S}^{(a)}_{11} + \mathbf{S}^{(a)}_{12}}
\mathbf{S}^{(b)}_{21} \left(\mathbf{1} - \mathbf{S}^{(a)}_{22}
\mathbf{S}^{(b)}_{11}\right)^{-1}\mathbf{S}^{(a)}_{21}~~, 
\label{str21} \\
\mathbf{S}_{22}  = & \mathbf{S}^{(b)}_{22} + \mathbf{S}^{(b)}_{21}
\mathbf{S}^{(a)}_{22}\left(\mathbf{1} - \mathbf{S}^{(b)}_{11}
\mathbf{S}^{(a)}_{22}\right)^{-1}\mathbf{S}^{(b)}_{12}~~. 
\label{str22}
\end{eqnarray}
\end{subequations}
Because of the presence of forward propagating modes only, the evaluation 
of the scattering matrix is numerically stable also for ``large'' systems.
Moreover, the composition rule (\ref{starp}) naturally accommodates a
different number of modes in the lead and in the cavity. These features
are of particular relevance in the present instance, where the stub's
width $c$ may vary over a rather large range of values \cite{noi07}.

It is worth to stress here that each block ${\mathbf S}_{ij}$ in the 
scattering operator ${\mathbf S}$ is itself a matrix, whose elements
are labeled by mode or channel indexes. For an incoming wave of unit
flux impinging from the left, $\left({\mathbf S}_{11}\right)_{nm}$ 
represents the reflection coefficient towards the left from the initial
channel $m$ into the final one $n$, whereas 
$\left({\mathbf S}_{21}\right)_{nm}$ is the transmission coefficient
to the right from mode $m$ into mode $n$. Similarly, 
$\left({\mathbf S}_{12}\right)_{nm}$ and $\left({\mathbf S}_{22}\right)_{nm}$
are the $m \rightarrow n$ transmission amplitudes to the left and reflection
coefficient to the right for an electron incoming from the right.
Once the transmission coefficients are known, the total conductance 
$G$ (in units $2e^2/h$) is given by the B\"uttiker formula 
\cite{da95,fg99,bi85}
\begin{equation}
G = \sum_{m,n} \frac{k_n^{(l)}}{k_m^{(l)}}\left\vert\left({\mathbf S}_{21}
\right)_{nm}\right\vert^2~~,
\label{but}
\end{equation}
where $k_n^{(l)}$ and $k_m^{(l)}$ denote the lead propagation momenta in 
channel $n$ and $m$, respectively, and the sum is restricted to the open 
channels in the duct.
 
The above $S$--matrix approach can be straightforwardly extended to complex
energies. We used our code to numerically locate the poles
of the $S$--matrix in the multi--sheeted energy surface. In the
following, sheets will be specified according to the sign of the
imaginary part of the lead momenta in the various channels \cite{lw67}.
Thus, for a four--channel situation, the physical sheet, where all
the imaginary parts of the momenta are positive, will be denoted
as $(++++)$, whereas on the sheet $(-+++)$ one has ${\rm Im}k^{(l)}_1
<0$ and ${\rm Im}k^{(l)}_i > 0$ for the other three channels. Dominant
poles producing resonance effects in the lowest subband, between
the first and second scattering thresholds $E_{T}^{(1)}$ and $E_{T}^{(2)}$,
are in the fourth quadrant of this sheet near the real energy axis,
and have $E_{T}^{(1)} \leq {\rm Re}E_p \leq E_{T}^{(2)}$.
Similarly, dominant poles for resonances in the second subband
with $E_{T}^{(2)} \leq E \leq E_{T}^{(3)}$ lie in sheet $(--++)$
and have $E_{T}^{(2)} \leq {\rm Re}E_p \leq E_{T}^{(3)}$.
We have chosen the value $m^\ast = 0.067 m_e$ for the effective electron
mass, which is appropriate for the Al$_x$Ga$_{1-x}$As/GaAs interface.
We verified that convergence is attained for both the conductance and the
pole positions when four channels are included in the external duct, and up to
ten channels are taken into account in the cavity. In these conditions,
the position of the poles in the complex energy plane can be guaranteed
with an accuracy of the order $10^{-5}$. From now on, to exploit
the scale invariance of the system, all lengths are measured in terms
of the waveguide width $b$, and energies in terms of the waveguide
fundamental mode $\epsilon^{(l)}_1 = \frac{\hbar^2}{2m^\ast}
\left(\frac{\pi}{b}\right)^2$, and the ``tilde'' symbol will be
used to denote adimensional quantities, so that one has for the various
thresholds $\tilde E^{(n)}_{T} = 1,\;2,\;\ldots$. The calculations we
present refer to a device with $\tilde l_s = 1$; a repulsive, double
Gaussian defect
\begin{equation}
\tilde V(\tilde x, \tilde y) \equiv \tilde V_0e^{-\tilde 
\beta^2(\tilde x - \tilde x_c)^2 - \tilde \alpha^2(\tilde y - \tilde y_c)^2}~
\label{dbg}
\end{equation}
centered in $(\tilde x_c, \tilde y_c) = (0.50,0.25)$ has been allowed
in the cavity. The decay constants along the transverse and propagation 
direction have been fixed at $\tilde \alpha = 15$, $\tilde \beta = 10$, so as 
to ensure that the potential is entirely contained within a region 
$\tilde w = 0.3$ wide and $\tilde l_d = 1$ long, displaced a distance 
$\tilde Y_0 = 0.1$ from the lower edge of the guide. The smooth dependence of 
$\tilde V(\tilde x, \tilde y)$ has been taken into account through a slicing 
technique, {\em i.e.}, replacing the actual interaction with a sequence of 
pseudodefects having a constant value along the $x$ direction 
\cite{noi07,sx96}. Quite stable results are obtained with $N=10\div 15$ slices.
In the present calculations we have chosen $\tilde V_0 = 4$.
\begin{figure}[ht]
\centerline{\includegraphics[width=10 truecm,angle=-90]{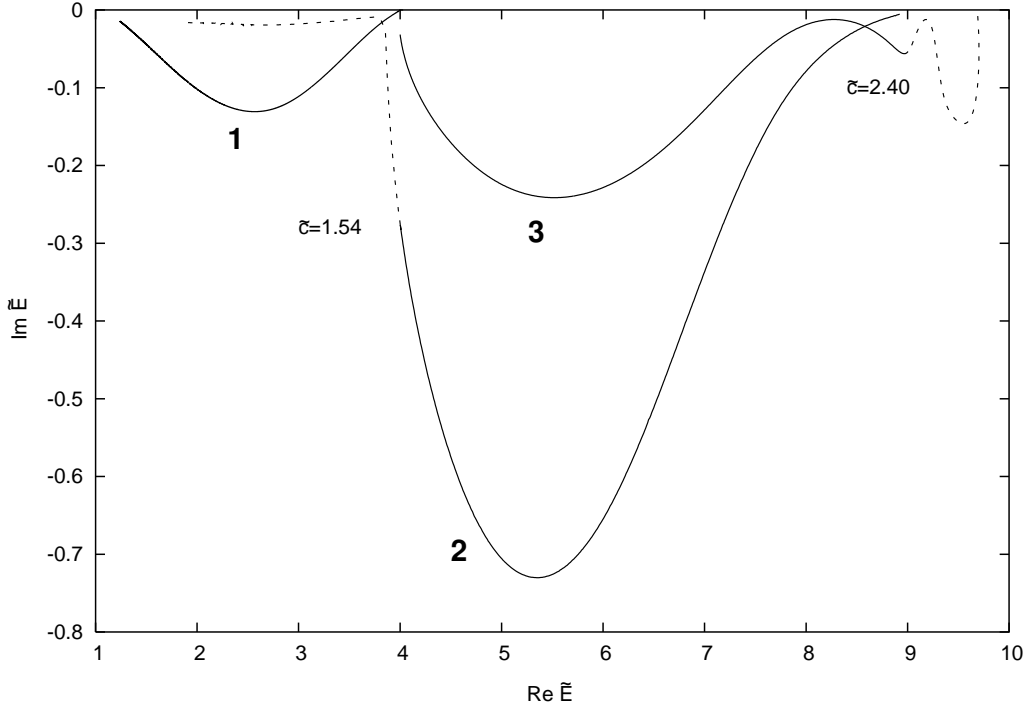}}
\caption{Motion of three $S$--matrix poles on the Riemann energy surface
with varying $\tilde c$. The three trajectories correspond to $1.50 \leq \tilde
c \leq 5.00$, $1.00 \leq \tilde c \leq 2.00$, and $1.33 \leq \tilde c \leq 
5.00$ for pole 1, 2, and 3, respectively. Note that pole 1 moves on the
$(-+++)$ sheet, whereas poles 2 and 3 belong to the $(--++)$ sheet. Shadow
and dominant poles are drawn as dashed and full lines, respectively. The
values of $\tilde c$ where a pole changes its nature are given in the figure.}
\label{fig2}
\end{figure}

In Fig. \ref{fig2} we report the trajectories on the complex energy surface
of three $S$--matrix poles with varying stub's width $\tilde c$. Pole 1
moves from the upper edge $\tilde E_{T}^{(2)}$ towards the lower edge 
$\tilde E_{T}^{(1)}$ of the first subband as $\tilde c$ is increased from 
$1.50$ to $5.00$. Similarly, pole 2 moves downwards from the third threshold 
passing below the second one as $\tilde c$ is increased from $\tilde c = 1.00$
up to $\tilde c = 2.00$, whereas pole 3 refer to $1.33 \leq \tilde c \leq 
5.00$. In all cases one has the ``binding'' effect typical of an increase of 
the stub's width \cite{noi07}. Note that the three pole trajectories appear
to be close to each other, but are in fact on different sheets of the
energy Riemann surface. Pole 1 lies on the $(-+++)$ sheet, and can
produce resonance effects in the first subband, whereas poles 2 and 3
belong to the $(--++)$ sheet, and are responsible of resonance structures
in the second subband. As a consequence, pole 2 is a dominant pole
until is passes below $E_{T}^{(2)}$, which happens for $\tilde c = 1.54$;
for $\tilde c > 1.54$ it becomes a shadow pole, since the $(--++)$
sheet is more distant from the first subband than the $(-+++)$
sheet, where the relevant resonance poles may be found. Similarly, 
pole 3 is shadow for $\tilde c < 2.40$, and becomes a dominant 
pole for greater values of $\tilde c$. In Fig. \ref{fig2} dominant
and shadow poles are drawn as full and dashed lines, respectively.

\begin{figure}[ht] 
\centerline{\includegraphics[width=10 truecm,angle=-90]{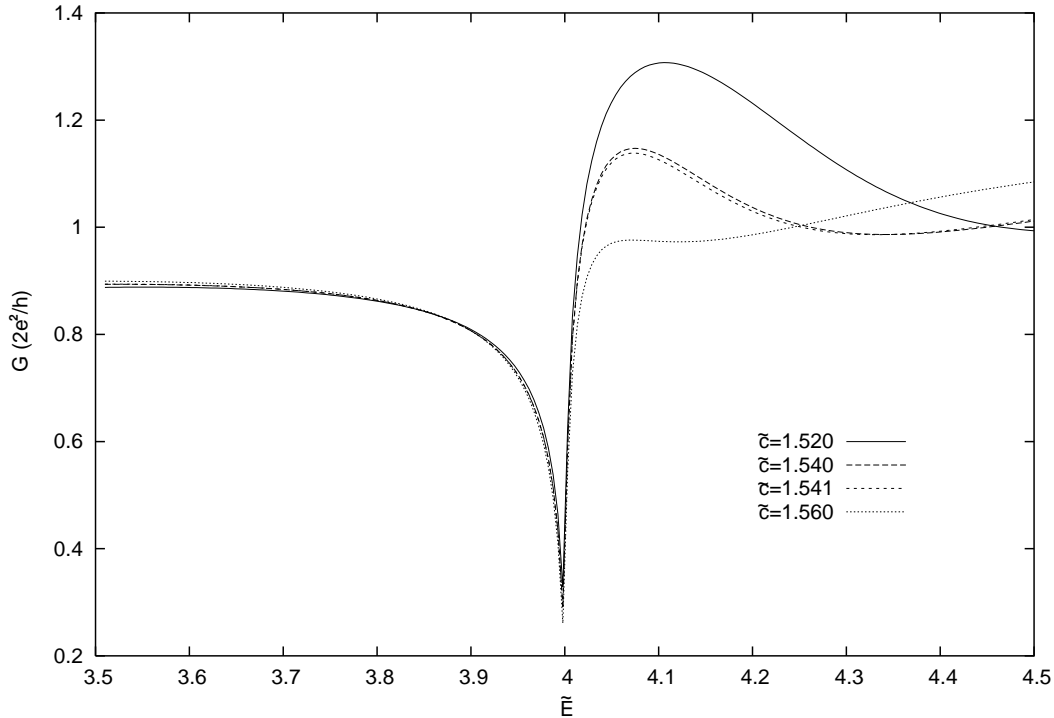}}
\caption{Conductance (in units $2e^2/h$) in the region of the second threshold 
for $\tilde c = 1.520$ (solid line), $\tilde c = 1.540$ (dashed line), 
$\tilde c = 1.541$ (short--dashed line), and $\tilde c = 1.560$ (dotted line).}
\label{fig3}
\end{figure}

The change of status of a pole from dominant to shadow pole as
it passes a threshold can explain the remarkable effects that even small
variations of $\tilde c$ may have on the conductance near threshold.
This is illustrated in Fig. \ref{fig3}, where we plot the conductance
in the second threshold region $(3.5 \leq \tilde E \leq 4.5)$, in 
correspondence to $\tilde c = 1.520,\;1.540,\;1.541,\;{\rm and}\; 1.560$. The
corresponding conductance profiles are given by the solid, long--dashed, 
short-dashed, and dotted lines, respectively. 

For $\tilde c = 1.520$ pole 2 is dominant, since one has $\tilde E_p \simeq 
4.14-0.39 i$, and produces the resonance peak one observes just above 
threshold. For $\tilde c = 1.540$ and $\tilde c = 1.541$ pole 2 is just above 
$(\tilde E_p \simeq 4.002-0.276i)$ and just below $(\tilde E_p \simeq 3.990 
- 0.270 i)$ threshold. One has that the resonance peak is still visible in 
both cases, which means that the dominant $\rightarrow$ shadow transition
does not prevent the pole from having effects on the observable quantities.
For $\tilde c = 1.560$ the pole has moved down to $\tilde E_p 
\simeq 3.87-0.09 i$, and it is far away enough from the physical region, to 
have no effects on the conductance, which appears rather flat above threshold. 
It is worth to stress that for these values of $\tilde c$ pole 1 is
far above the second threshold, and cannot influence the conductance profile 
in the first subband; as a matter of fact, in all cases the conductance is 
practically the same below threshold, and exhibits a cusp structure, with 
infinite slope as a function of energy both from above and from below.
This behavior is indeed discernible at threshold in all calculations, and
can be explained much in the same way, as one explains threshold phenomena
in inelastic scattering processes. When a new transverse mode opens up, less
energy is available in the propagation direction, so that one has the analogue
of ``endoergic'' reactions in inelastic scattering \cite{rn82}. From Eq.
(\ref{but}) one sees that $G$ is linear with respect to the corresponding
final momentum $k_n^{(l)}$. Since  $k_n^{(l)}$ is related to the total
energy $E$ and to the relevant waveguide eigenenergy $\epsilon^{(l)}_n$
by \cite{noi07}
\begin{equation}
k_n^{(l)} = \left[2m^\ast\left(E - \epsilon^{(l)}_n \right)/\hbar^2
\right]^{1/2}~,
\nonumber
\end{equation}
one actually expects an infinite derivative of $G$ with respect to $E$ 
\cite{rn82}.      

The effects due to the exchange of role between shadow and dominant
poles are illustrated in Fig. \ref{fig3}, where we plot $G$ near the
second threshold for $\tilde c = 1.54$ (solid line) and $\tilde c = 1.70$
(dashed line). In the former case one has the resonance peak above
threshold due to pole 2, as discussed previously; in the latter, pole 2 has 
moved down to $\tilde E_p \simeq 2.68 - 0.02 i$ and has no effect on the 
conductance any longer; pole 1 which moves on the $(-+++)$ sheet, on the other 
hand, is now dominant, being located at $\tilde E_p \simeq 3.78 - 0.02 i$, and
produces the Fano dip one observes in Fig. \ref{fig3}. Note that in the
first subband one can have the simultaneous presence of poles and transmission 
zeros, which cannot occur when more than a propagating mode are active.
 
\begin{figure}[ht] 
\centerline{\includegraphics[width=10 truecm,angle=-90]{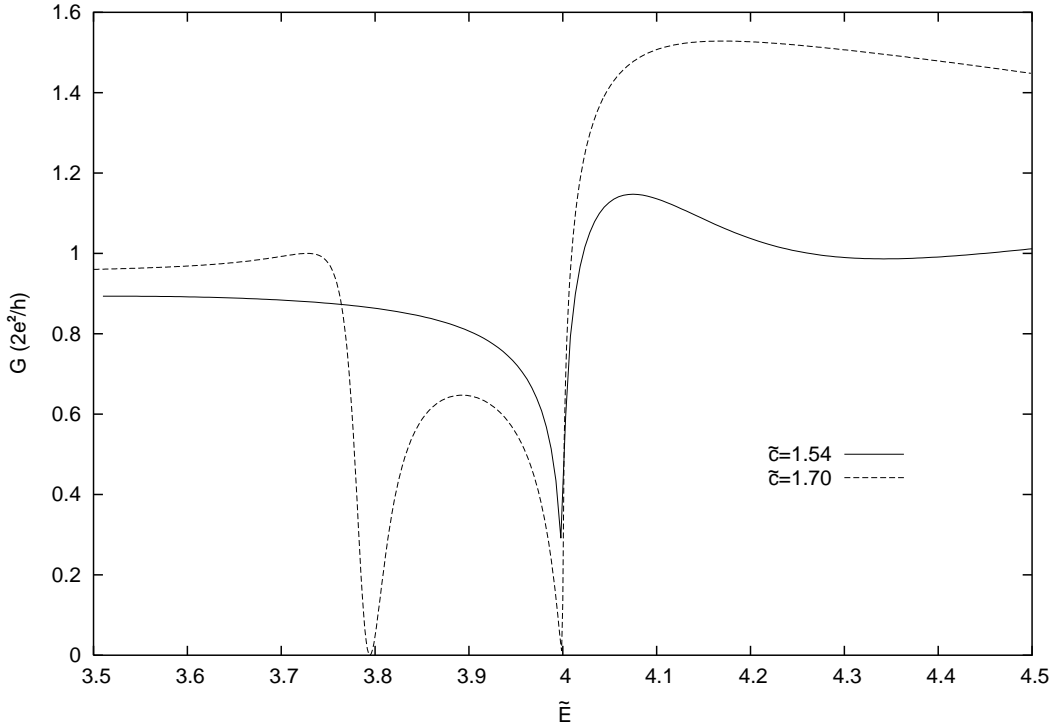}}
\caption{Conductance (in units $2e^2/h$) in the region of the second threshold
for $\tilde c = 1.54$ (solid line) and $\tilde c = 1.70$ (dashed line).}
\label{fig4}
\end{figure}

\begin{figure}[ht] 
\centerline{\includegraphics[width=10 truecm,angle=-90]{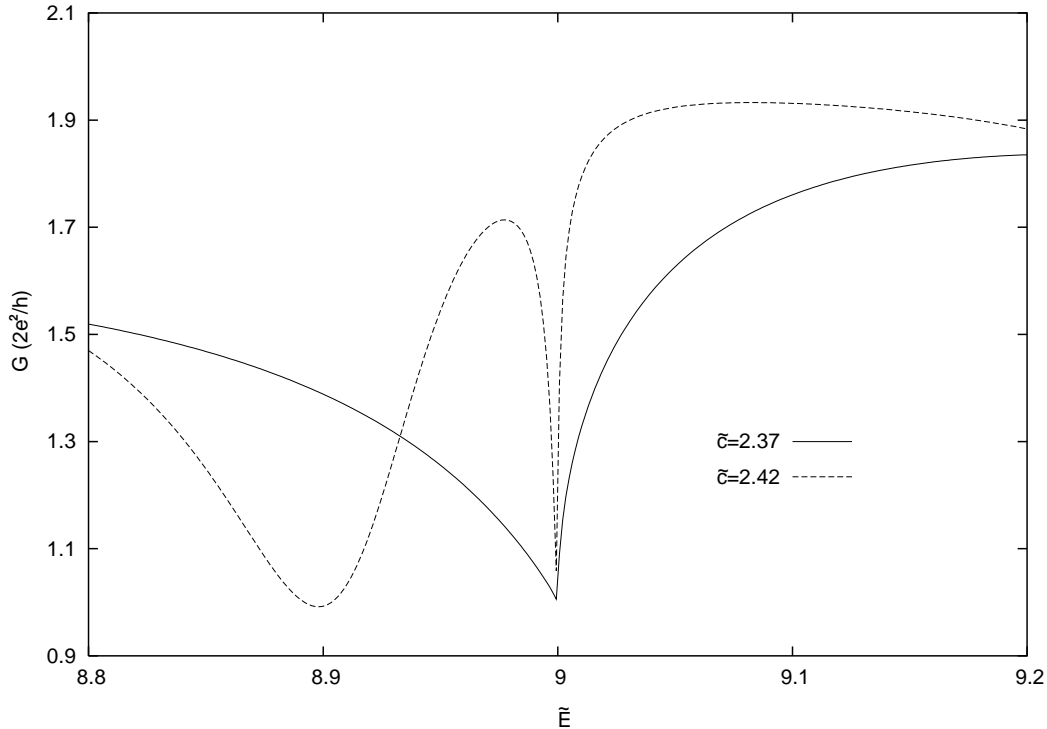}}
\caption{Conductance (in units $2e^2/h$) in the region of the third threshold
for $\tilde c = 2.37$ and $\tilde c = 2.42$.}
\label{fig5}
\end{figure}

A similar phenomenon is visible in correspondence to the third threshold.
An example is given in Fig. \ref{fig5}, where the conductance
around the third threshold is plotted for $\tilde c = 2.37$ and $\tilde c 
= 2.42$. While a resonance dip is clearly visible in the latter case, no 
resonance at all is discernible for the shorter stub, and only the threshold 
cusp survives for the conductance profile. Such a striking change in
correspondence to so small a change in the cavity width can be readily
explained in terms of a dominant to shadow pole transition. Indeed,
for $\tilde c = 2.42$ pole 3 of Fig. \ref{fig2} is located at 
$\tilde E_p \simeq 8.92 - 0.053 i$ in the $(--++)$ sheet, and plays
the role of dominant pole for the second subband. When the stub
is shortened, the pole moves on its sheet up to $\tilde E_p \simeq 9.11 - 
0.023 i$, in correspondence to the third subband, and becomes a shadow 
pole.

In conclusion, we have demonstrated that the behavior of the conductance
near the thresholds for the opening of new propagating modes, and its
sometimes striking changes in correspondence to moderate or even small
variations of the stub's width are signals of the transition from a
dominant to a shadow status of the $S$--matrix poles. This
result shows that concepts and methods of the analytic $S$--matrix, 
widely employed in traditional scattering theory, may have their
counterpart in the analysis of systems with a confined geometry.

\end{document}